\documentclass[twocolumn,showpacs,prb]{revtex4}

\usepackage{graphicx}
\usepackage{amsmath}
\usepackage{color}
\usepackage{hyperref}

\renewcommand{\Im}{\mathop{\rm Im}}

\newcommand{\Tr}{\mathop{\rm Tr}}
\newcommand{\eps}{\varepsilon}
\newcommand{\sgn}{\mathop{\rm sgn}}

\begin{document}

\title{The anisotropic conductivity of two-dimensional electrons on a half-filled high Landau level}

\author{I.\,S.\,Burmistrov}

\affiliation{L.D. Landau Institute for Theoretical Physics,
Kosygina str. 2, 117940 Moscow, Russia}
\affiliation{Institute for
Theoretical Physics, University of Amsterdam, Valckenierstraat 65,
1018 XE Amsterdam, The Netherlands}

\begin{abstract}We study the conductivity of two-dimensional
interacting electrons on the half-filled $N$th Landau level with
$N \gg 1$ in the presence of the quenched disorder. The existence
of the unidirectional charge-density wave state at temperature
$T<T_c$, where $T_c$ is the transition temperature, leads to the
anisotropic conductivity tensor. We find that the leading
anisotropic corrections are proportional to $(T_c-T)/T_c$ just
below the transition in accordance with the experimental findings.
Above $T_c$ the correlations corresponding to the unidirectional
charge-density wave state below $T_c$ result in the corrections to
the conductivity proportional to $\sqrt{T_c/(T-T_c)}$.
\end{abstract}

\maketitle
%
\textbf{1. Introduction.} Two-dimensional electrons in a
perpendicular magnetic field was a subject of intensive studies,
both theoretical and experimental, for several
decades~\cite{AFS,QHE}. It has been found that the properties of
two-dimensional electrons in the magnetic field are strongly
affected by the presence of electron-electron interaction as well
as by impurities. The behaviour of the system in a strong magnetic
field where only the lowest Landau level is occupied has been
investigated in great details~\cite{QHE}. But only several
attempts were made to consider the system in a \textit{weak}
magnetic field (large number of Landau levels $N \gg 1$ are
occupied) where the Coulomb energy at distances of the order of
the magnetic length exceeds the cyclotron energy~\cite{Attempts}.

\begin{figure}[tbp]
\includegraphics[width=120pt]{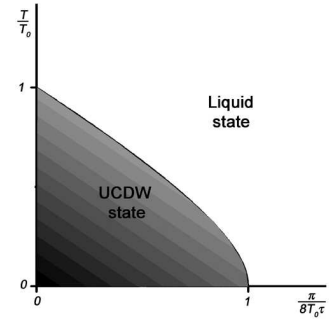}
\caption{Phase diagram on a half filled high Landau level.}
\label{FIG2}
\end{figure}

The progress in understanding the clean two-dimensional electrons
in a weak magnetic field was achieved by Aleiner and Glazman who,
by using the small parameter $1/N \ll 1$, have derived the
successive theory that describes electrons on the partially filled
$N$th Landau level ~\cite{AG}. By treating the effective
electron-electron
 interaction on the $N$th Landau level within the Hartree-Fock
approximation, Koulakov, Fogler, and Shklovskii~\cite{KFS}
predicted a unidirectional charge-density-wave (UCDW) state
(stripe phase) for the half-filled high Landau level at zero
temperature and in the absence of disorder. Moessner and
Chalker~\cite{MC} showed the existence of the UCDW state on the
half-filled high Landau level without disorder below some
temperature $T_{0}$. In the presence of disorder the
 UCDW state on the half-filled high Landau level can
exist  if the Landau level broadening $1/2\tau$ does not exceed
the critical value $1/2\tau_c=4 T_0/\pi$~\cite{Burm1} (see
Fig.~\ref{FIG2}). (We use the system of units with $\hbar=1$,
$c=1$ and $k_B=1$ throughout the Letter.)

The anisotropic magnetoresistance discovered near half-fillings of
Landau levels at low temperatures was attributed to the existence
of the UCDW state ~\cite{Exp}. This stimulates an extensive study
of the properties of two-dimensional electrons in a weak magnetic
field~\cite{Fogler}. However up to date, the magnetoresistance of
the UCDW state have been theoretically considered in the zero
temperature limit only where stripes have well-defined
edges~\cite{Fogler}.

The main objective of the present Letter is to present the results
for the conductivity tensor of the UCDW state developed on the
half-filled high Landau level in the presence of the quenched
disorder just below the transition temperature $T_c$ where the
expansion in the CDW order parameter $\Delta$ is justified.

%
\textbf{2. UCDW state.} The two-dimensional electrons in a weak
perpendicular magnetic field $H$ occupy large number ($N \gg 1$)
of the Landau levels. We assume that disorder is \textit{weak} so
it leads to the Landau level broadening $1/2\tau$ that satisfies
the condition $1/2\tau \ll \omega_H$, where $\omega_H=e H/m$ is
the cyclotron frequency with $e$ and $m$ being the electron charge
and the effective electron mass respectively. As the temperature
decreases the second-order transition from the homogeneous state
to the UCDW state occurs (see Fig.~\ref{FIG2}). Vector
$\textbf{\textit Q}$ that characterizes a period of the UCDW can
be oriented along spontaneously chosen direction. Usually, the
orientation is fixed either by the intrinsic anisotropy of the
crystal or by the small external in-plane magnetic
field~\cite{Fogler}. Hereinafter, we assume that the vector
$\textbf{\textit Q}$ is directed under angle $\phi$ with respect
to the $x$ axis. The period of the UCDW is seemed to be of the
order of the cyclotron radius $R_c= l_H \sqrt{2 N+1}$, where
$l_H=1/\sqrt{m \omega_H}$ denotes the magnetic length. More
precisely, the modulus of the vector $\textbf{\textit Q}$ equals
$Q=r_0/R_c$, where $r_0\approx 2.4$ is the first zero of the
zeroth order Bessel function of the first kind
$\mathcal{J}_0(z)$~\cite{KFS,Burm1}.

The temperature $T_c$ of the second-order transition from the
homogeneous to UCDW state is determined as the solution of the
following equation~\cite{Burm1},
\begin{equation}\label{SOT}
\frac{T_c}{T_0} = \frac{2}{\pi^2}\zeta\left (2, \frac{1}{2}
+\frac{1}{4 \pi T_c \tau}\right ),
\end{equation}
where $\zeta(2,z)$ is the generalized Riemann zeta function and
$T_0$ the transition temperature in the \textit{clean} case
($1/\tau =0$). We notice that Eq.\eqref{SOT} has the solution for
$T_c$ only if the Landau level broadening is smaller than the
critical one $1/2\tau \leq 1/2\tau_c=4 T_0/\pi$ as it is shown in
Fig.~\ref{FIG2}. According to Refs.~\cite{KFS,MC},
\begin{equation}
T_{0}=\frac{r_{s}\omega _{H}}{4\pi \sqrt{2}}\left[ \ln
\Bigl(1+\frac{c}{ r_{s} }\Bigr)-\frac{c}{\sqrt{2}+r_{s}}\right],
\quad \frac{1}{N}\ll r_s \ll 1 \label{T0rs}
\end{equation}
where $c=1/(\sqrt{2}r_{0})\approx 0.3$, and $r_s = \sqrt{2}
e^2/\varepsilon v_F \ll 1$ with $v_F$ and $\varepsilon$ being the
Fermi velocity and the dielectric constant of a media
respectively. It is worth mentioning that the $T_0$ is determined
by the characteristic energy $e^{2}/R_{c}\sim r_{s}\omega _{H} \ll
\omega _{H}$ of the screened electron-electron interaction on the
$N$th Landau level, cf. Eq.\eqref{U01}.

%

\textbf{3. Results.} The conductivity tensor $\sigma_{ab}$ of the
two-dimensional electrons on the half-filled high Landau level
above the transition temperature $T_c$, i.e. in the homogeneous
state, is known to be isotropic~\cite{AFS}. To this end we show
that in the presence of the \textit{quenched weak} disorder the
existence of the UCDW state on the half-filled high Landau level
below $T_c$ results in the \emph{anisotropic} corrections to the
isotropic conductivity tensor $\sigma_{ab}$. For the temperature
slightly below $T_c$, where the condition $T_c-T \ll T_c$ is hold,
the anisotropic corrections are given as
\begin{equation}
\left . \begin{array}{c}
\delta \sigma_{xx}^{(\rm anis)} \\
\delta \sigma_{yy}^{(\rm anis)}
\end{array}
\right \} = \mp N f\left (\eta_c\right )\mathcal{G}\left
(\eta_c\right ) \cos[2 \phi] \frac{T_c-T}{T_c},\label{cond1}
\end{equation}
and
\begin{equation}
 \left . \begin{array}{c}
\delta \sigma_{xy}^{(\rm anis)} \\
\delta \sigma_{yx}^{(\rm anis)}
\end{array}
\right \} = N f\left (\eta_c\right )\mathcal{G}\left (\eta_c\right
) \sin[2 \phi] \frac{T_c-T}{T_c}.\label{cond2}
\end{equation}
Here for convenience we introduce the dimensionless parameter
$\eta_c=1/4\pi T_c\tau$ that we use throughout the Letter. The
functions $f(z)$ and $g(z)$ are defined as
\begin{equation}\label{fn}
f(z) = \frac{32 \mathcal{J}_1^2(r_0)
z^3}{\left(\frac{1}{2}+z\right )^5},\,\, g(z) = \frac{16 \pi
z}{\left(\frac{1}{2}+z\right )\left (z^{2}+\left (
\frac{1}{2}+z\right )^2\right )},
\end{equation}
where $g(z)$ will be used below. The other function
$\mathcal{G}(z)$ is given as
\begin{equation}\label{G}
\mathcal{G}(z) = \frac{\zeta(2,\frac{1}{2}+z)- z \zeta(3,
\frac{1}{2}+z)}{-3 \zeta(4,\frac{1}{2}+z) + 4 \Phi_0(z)+2
\Phi_2(z)},
\end{equation}
with
\begin{equation}
\Phi_{n}(z) =\frac{\zeta( 2,\frac{1}{2}+z)}{z^{2}\mathcal{J}
_{0}^{2}(n r_{0})} -\frac{\Im \psi( \frac{1}{2}+z+i\,z
\mathcal{J}_{0}(nr_{0}))}{z^{3}\mathcal{J} _{0}^{3}(n r_{0})}.
\label{Phi}
\end{equation}
The $\psi(z)$ stands for the digamma function and symbol $\Im$
denotes the imaginary part.

There are several features of the main results \eqref{cond1} and
\eqref{cond2}.  First of all, the anisotropic corrections
$\delta\sigma_{ab}^{(\rm anis)}$ are proportional to
$(T_c-T)/T_c$. Although, the Eqs.\eqref{cond1} and \eqref{cond2}
are derived only for the case of a short-range random potential
(quenched disorder), it can be shown that the anisotropic
corrections remain proportional to $(T_c-T)/T_c$ in the case of a
long-range random potential as well~\cite{Burm3}. We emphasize
that such temperature dependence of the developing anisotropy in
magnetoresistance was observed in the experiments~\cite{Exp}.

The angle dependence of the anisotropic correction \eqref{cond1}
to conductivity $\sigma_{xx}$ has the minimum for $\phi=0$, that
corresponds to the vector $\textbf{\textit Q}$ directed along the
$x$ axis and stripe oriented along the $y$ axis. From
Eq.\eqref{cond1} we see that the conductivity $\sigma_{yy}$ along
the stripe
 is enhanced
whereas the conductivity $\sigma_{xx}$ across the stripes (along
the modulation of the order parameter) is suppressed as it should
be according to the experiments~\cite{Exp}. In the same time, the
anisotropic correction \eqref{cond2} to $\sigma_{xy}$ vanishes. If
the vector $\textbf{\textit Q}$ is oriented at angle $\phi=\pi/4$
with respect to the $x$ axis, the anisotropic correction
\eqref{cond1} to $\sigma_{xx}$ becomes zero due to the symmetry
between $x$ and $y$ axes. Conversely, the anisotropic correction
\eqref{cond2} to $\sigma_{xy}$ is attained the minimum .

\begin{figure}[t]
\centerline{\includegraphics[width=200pt,height=150pt]{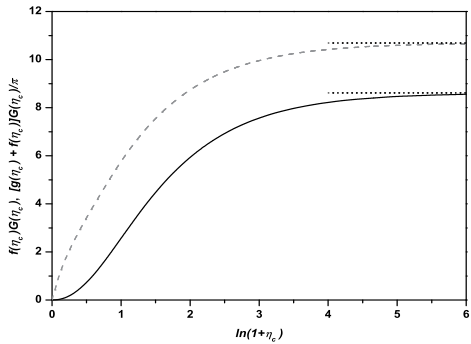}}
\caption{The $f(\eta_c)\mathcal{G}(\eta_c)$ and
$[g(\eta_c)+f(\eta_c)]\mathcal{G}(\eta_c)/\pi$ as functions of
$\ln(1+\eta_c)$.} \label{FIG3}
\end{figure}

The behavior of the anisotropic corrections \eqref{cond1} and
\eqref{cond2} as the functions of the parameter $\eta_c$ for fixed
temperature $T$ and angle $\phi$ are shown in Fig.~\ref{FIG3}.

In addition, the existence of the UCDW state on the half-filled
high Landau level leads to the isotropic correction that for $T_c-
T\ll T_c$ is as follows
\begin{equation}\label{cond3}
\delta \sigma_{xx}^{(\rm isot)} = - \frac{N}{\pi}\left [ g\left
(\eta_c\right )+ f\left (\eta_c\right )\right ] \mathcal{G}\left
(\eta_c\right )\frac{T_c-T}{T_c}.
\end{equation}
The behavior of the isotropic correction \eqref{cond3} as the
functions of $\eta_c$ for fixed temperature $T$ is shown in
Fig.~\ref{FIG3}.

%
{\bf 4. Model.} The grand canonical partition function of the
two-dimensional interacting electrons in the random potential
$V(\textit{\textbf{\textit r}})$ subjected to the perpendicular
constant magnetic field $H$ and the time-dependent external vector
potential $\textbf{\textit A}$ is given by
\begin{equation}
\mathcal{Z} =\int \mathcal{D}[\Psi,\Psi^\dagger]
\mathcal{D}[V]\,\mathcal{P}[V]\,\exp \mathcal{S}_0[\Psi
,\Psi^\dagger,V], \label{ZStart0}
\end{equation}
where the action $\mathcal{S}_0[\Psi,\Psi^\dagger,V]$ in the
Matsubara representation has the form
\begin{gather}
\mathcal{S}_0=\int\limits_{r}\Psi^{\dagger}(\textit{\textbf{\textit
r}}) \Bigl[i\omega+\mu -\mathcal{H}_0+ \hat{K}-V(
\textit{\textbf{\textit r}})\Bigr]\Psi(\textit{\textbf{\textit
r}})\label{Sinit0}
\\
-\frac{T}{2}\int\limits_{r r^{\prime }}\sum_{\alpha\nu_n}
\Psi^{\dagger}(\textit{\textbf{\textit r}})
I^{\alpha}_n\Psi(\textit{\textbf{\textit r}})
\frac{e^2}{\varepsilon|\textit{\textbf{\textit
r}}-\textit{\textbf{\textit r}}^{\prime}|}
\Psi^{\dagger}(\textit{\textbf{\textit
r}}^{\prime})I^{\alpha}_{-n}\Psi (\textit{\textbf{\textit
r}}^{\prime}). \notag
\end{gather}
Here  we use the matrix notation $\Psi^\dagger (\cdots)\Psi =
\Psi^{\dagger\alpha }_{\omega_{n}}
(\cdots)^{\alpha\beta}_{nm}\Psi^{\beta}_{\omega_{m}}$ for the
electron annihilation $\Psi^{\alpha
}_{\omega_{n}}(\textit{\textbf{\textit r}})$ and creation
$\Psi^{\dagger\alpha }_{\omega_{n}}(\textit{\textbf{\textit r}})$
operators. Superscripts $\alpha,\beta=1,\cdots,2 N_r$ stands for
replica indices combined with spin ones. We introduce the replica
indices in order to average over the random potential
$V(\textit{\textbf{\textit r}})$. The subscripts
$\omega_n,\omega_m$ denote the Matsubara fermionic frequencies
$\omega_n=\pi T(2n+1)$. The one-particle hamiltonian
$\mathcal{H}_0$ for a two-dimensional electron in the presence of
the magnetic field $H=\epsilon _{ab}\partial _{a}A^{\rm st}_{b}$
is defined as $ \mathcal{H}_0=-\textit{\textbf{\textit D}}^2/(2m)$
with the covariant derivative $\textit{\textbf{\textit D}} =
\nabla- i e \textit{\textbf{\textit A}}^{\rm st}$. The matrix
$\omega$ has only diagonal elements
$\omega_{nm}^{\alpha\beta}=\omega_n
\delta^{\alpha\beta}\delta_{nm}$ that represent the Matsubara
frequencies $\omega_n$. The matrix $I^\alpha_n$ with elements
$(I^{\alpha}_{n})^{\beta\gamma}_{lm}=
\delta^{\beta\alpha}\delta^{\gamma\alpha}\delta_{m-l,n}$ is a
generator of the $U(1)$ gauge transformation. The time-dependent
external vector potential $\textit{\textbf{\textit A}}$ is
involved through the matrix $\hat K = \sum_{\alpha,n} K(\nu_n)
I^{\alpha}_n$ where
\begin{equation}\label{K0}
K(\nu_n) = -\frac{e}{m}\textit{\textbf{\textit A}}(\nu_n)
\textit{\textbf{\textit D}} +\frac{e^2}{2m} \sum_{\nu_m}
\textit{\textbf{\textit A}}(\nu_{n-m})\textit{ \textbf{\textit
A}}(\nu_m).
\end{equation}
Here $\textit{\textbf{\textit A}}(\nu_n)$ is the Fourier component
of the external vector potential $\textit{\textbf{\textit A}}$
with frequency $\nu_n=2\pi T n$.

We assume the white-noise distribution for the random potential
$V(\textit{\textbf{\textit r}})$
\begin{equation}
\mathcal{P}[V(\textit{\textbf{\textit r}})]=\frac{1}{\sqrt{2\pi
g}}\,\exp \left[ -\frac{1}{2
g}\int_{r}V^{2}(\textit{\textbf{\textit r}})\right]. \label{dis}
\end{equation}
This distribution corresponds to a short-range random potential
with the correlation length $d$ smaller than the magnetic field
length $l_{H}$, $d \ll l_H$. In high mobility samples used in
experiments~\cite{Exp}, however, the disorder potential has
long-range correlations. In the case $d \gg l_H$ one should
distinguish between the Landau level broadening $1/2\tau$ and the
inverse transport time $1/\tau_{\rm tr}$. Therefore, the
anisotropic \eqref{cond1} and \eqref{cond2} as well as isotropic
\eqref{cond3} contributions will be determined by both energy
scales $1/\tau$ and $1/\tau_{\rm tr}$ those ratio depends on the
value of dimensionless parameter $d/l_H$~\cite{AFS}. Nevertheless,
the main result that the anisotropic as well as isotropic
corrections to $\sigma_{ab}$ are proportional to $(T_c-T)/T_c$
will survive for a long-range random potential~\cite{Burm3}.

%
\textbf{5. Method.} To proceed we integrate over the random
potential $V(\textit{\textbf{\textit r}})$ in Eq.\eqref{ZStart0}.
As usual,  it leads to the quartic interaction that we decouple by
introducing the matrix field $Q(\textit{\textbf{\textit
r}})$~\cite{ELK}. The annihilation $\Psi(\textit{\textbf{\textit
r}})$ and creation $\Psi^{\dagger}(\textit{\textbf{\textit r}})$
operators written in the basis of the eigenfunctions
$\phi_{pk}(\textit{\textbf{\textit r}})$ of the Hamiltonian
$\mathcal{H}_0$
\begin{equation}\label{expansion}
\Psi(\textit{\textbf{\textit r}}) = \sum_{p,k}
\psi_{pk}\phi_{pk}(\textit{\textbf{\textit r}}),\,\,
\Psi^{\dagger}(\textit{\textbf{\textit r}}) = \sum_{p,k}
\psi_{pk}^{\dagger}\phi_{pk}^{*}(\textit{\textbf{\textit r}})
\end{equation}
involve the electron states on all Landau levels. Therefore, the
term with the electron-electron interaction in the action
\eqref{Sinit0} contains the interactions of electrons from
different Landau levels. In general, to treat the problem
\eqref{Sinit0} analytically seems to be impossible. However, as it
was shown in Ref.~\cite{AG}, if the $N-1$ Landau levels are filled
whereas the $N$th Landau level is partially occupied, one can
obtain the description of the system in terms of electrons on the
$N$th Landau level only provided that the relative strength of the
bare Coulomb interaction is \textit{small}, $r_s \ll 1$, and the
magnetic field is rather \textit{weak}, $N r_s \gg 1$.

Following the same strategy as in Ref.~\cite{Burm2}, we obtain the
grand canonical partition function $\mathcal{Z}$ as
\begin{equation}
\mathcal{Z} =\int \mathcal{D}[\Psi,\Psi^\dagger]
\mathcal{D}[Q]\exp \mathcal{S}[\Psi ,\Psi^\dagger,Q],
\label{ZStart}
\end{equation}
where
\begin{gather}
\mathcal{S}=\int\limits_{r}\Psi^{\dagger}(\textit{\textbf{\textit
r}}) \Bigl[i\omega+\mu -\mathcal{H}_0+ \hat{K}+i
Q\Bigr]\Psi(\textit{\textbf{\textit r}}) - \frac{1}{2g}\Tr Q^2
\notag \\
-\frac{T}{2}\int\limits_{r r^{\prime }}\sum_{\alpha\nu_n}
\psi^{\dagger}(\textit{\textbf{\textit r}})
I^{\alpha}_n\psi(\textit{\textbf{\textit r}}) U_{\rm scr}(
\textit{\textbf{\textit r}},\textit{\textbf{\textit r}}^{\prime})
\psi^{\dagger}(\textit{\textbf{\textit
r}}^{\prime})I^{\alpha}_{-n}\psi (\textit{\textbf{\textit
r}}^{\prime}).\label{Sinit}
\end{gather}
Here symbol $\Tr$ denotes the trace over the Matsubara, replica
combined with spin and spatial indices. The electron-electron
interaction is written in terms of electron operator
$\psi(\textit{\textbf{\textit r}})= \sum_{k}
\psi_{Nk}\phi_{Nk}(\textit{\textbf{\textit r}})$ on the $N$th
Landau level only. The screened interaction $U_{\rm
scr}(\textit{\textbf{\textit r}})$ of electrons on the $N$th
Landau level takes into account the effects of electrons on the
other levels and has the following form ~\cite{AG,Burm2}
\begin{equation}
U_{\rm scr}(q)=\frac{2\pi e^{2}}{\varepsilon
q}\frac{1}{1+\displaystyle \frac{2\left
(1-\frac{\pi}{6\omega_H\tau}\right )}{ qa_{B}}\left( 1-\mathcal{J}
_{0}^{2}(q R_{c})\right)}. \label{U01}
\end{equation}
It is worth mentioning that the range of the screened
electron-electron interaction (\ref{U01}) is determined by the
Bohr radius $a_{B}=\eps/m e^2$. We assume that the magnetic field
is so weak that the condition $N r_s^2\gg 1$ is hold. That means
that the range $a_B$ of the screened electron-electron interaction
\eqref{U01} is much less than the magnetic length $l_H$. It allows
us to treat the interaction in the Hartree-Fock
approximation~\cite{MC}.

In the absence of the external vector potential
$\textit{\textbf{\textit A}}$ one can project the first line in
Eq.\eqref{Sinit} onto the $N$th Landau level, i.e. substitute
$\Psi(\textit{\textbf{\textit r}})\to \psi(\textit{\textbf{\textit
r}})$. Then the action becomes to involve electrons on the $N$th
Landau level only and, evidently, it simplifies the analysis. The
accuracy of such projection is of the order of $\max
\{T,\tau^{-1}\}/\omega_H \ll 1$. It is worthwhile to mention that
the correction $\pi/6\omega_H\tau\ll 1$ in the screened
interaction \eqref{U01} results in the correction of the same
order to the $T_0$. For reasons to be explained shortly we neglect
this effect. However, in order to investigate the response of the
system to the external vector potential $\textit{\textbf{\textit
A}}$ such projection onto the $N$th Landau level is not
appropriate. We should leave the action \eqref{Sinit} as it stands
because the matrix elements $D^a_{p_1p_2} = \int_r
\phi_{p_1k}^{*}(\textit{\textbf{\textit r}})D^a
\phi_{p_2k}(\textit{\textbf{\textit r}})$ of the covariant
derivative $\textit{\textbf{\textit D}}=(D^x,D^y)$ involve the
electron states on the adjacent Landau levels. As the last thing
we mention that the electrons on the $N$th Landau level should be
regarded as spin-polarized according to the numerical
findings~\cite{WS}.

The action \eqref{Sinit} involves the unitary matrix field
$Q(\textit{\textbf{\textit r}})$. There exists the saddle-point
solution $Q(\textit{\textbf{\textit r}}) = W^{-1} U_{\rm sp} W$ in
the absence of the electron-electron interaction. Here the
constant unitary matrix $W$ describes the global rotation, whereas
$(U_{\rm sp})^{\alpha\beta}_{nm} =\delta^{\alpha\beta}\delta_{nm}
\sgn \omega_{n}/2\tau$ with $1/2\tau = \sqrt{g/2\pi l_H^2}$. Being
motivated by the form of the saddle-point solution we split the
matrix field $Q(\textit{\textbf{\textit r}})$ in transverse
$W(\textit{\textbf{\textit r}})$ and longitudinal
$U(\textit{\textbf{\textit r}})$ components as
$Q(\textit{\textbf{\textit r}})=W^{-1}(\textit{\textbf{\textit
r}}) (U_{\rm sp}+U(\textit{\textbf{\textit
r}}))W(\textit{\textbf{\textit r}})$. As it is well-known the
transverse field $W(\textit{\textbf{\textit r}})$ is responsible
for weak localization corrections~\cite{ELK} but in the case of
interest they are of the order of $\ln N/N \ll 1$. Therefore we
eliminate the transverse field from the future considerations by
formally putting $W(\textit{\textbf{\textit r}})=1$. The
transformation of the variable $Q(\textit{\textbf{\textit r}})$
discussed above leads to the additional measure in the functional
integral~\cite{Pruisken}
\begin{equation}
\ln I[U] \sim \int\limits_r  \sum \limits_{\omega_n
\omega_m}^{\alpha \beta} \left [ 1 - \Theta(n m) \right ] U_{n
n}^{\alpha \alpha}(\textit{\textbf{\textit r}}) U_{m m}^{\beta
\beta}(\textit{\textbf{\textit r}}), \label{mesI}
\end{equation}
where $\Theta(x)$ stands for the Heaviside step function.

To describe the UCDW state we introduce the CDW order parameter
$\Delta$ that is related with a distortion of the electron density
on the $N$th Landau level
\begin{equation}\label{ed}
\langle \delta\rho(\textit{\textbf{\textit q}})\rangle = 2\pi
l^{-2}_H F_{NN}(\textit{\textbf{\textit q}})
[\delta(\textit{\textbf{\textit q}}-\textit{\textbf{\textit
Q}})+\delta(\textit{\textbf{\textit q}}+\textit{\textbf{\textit
Q}})]\Delta,
\end{equation}
where the $F_{p_1p_2}(\textit{\textbf{\textit q}})$ is defined as
\begin{equation}\label{H12}
F_{p_1p_2}(\textit{\textbf{\textit q}}) =2\pi l_H^2\sum\limits_k
\phi_{p_1 k}^{*}(0) \phi_{p_2 k}(\textit{\textbf{\textit q}}l_H^2)
e^{i \frac{1}{2}q_x q_y l_H^2}.
\end{equation}
In particularly, the form-factor $F_{NN}(q) \approx
\mathcal{J}_0(q R_c)$ for $q R_c \ll 2 N$. The presence of the
distortion of the electron density by the charge-density wave on
the $N$th Landau level results in the additional periodic
potential $\lambda(\textit{\textbf{\textit r}})$ that is related
with the UCDW order parameter as
\begin{equation}\label{lambda}
\lambda(\textit{\textbf{\textit q}}) = (4\pi)^2 T_0 F_{NN}^{-1}(q)
\Bigl[\delta (\textit{\textbf{\textit q}}-\textit{\textbf{\textit
Q}})+\delta (\textit{\textbf{\textit q}}+\textit{\textbf{\textit
Q}})\Bigr]\Delta.
\end{equation}

After the Hartree-Fock decoupling~\cite{FPA} of the interaction
term in the action \eqref{Sinit} and integration over electrons,
we obtain
\begin{equation}\label{Z1}
\mathcal{Z} \sim \int \mathcal{D}[U]I[U] \exp \mathcal{S}[U],
\end{equation}
where the action becomes
\begin{equation}\label{S1}
\mathcal{S} = - \frac{1}{2g}\Tr U^2 + \Tr\ln \left (1+ (i U + \hat
K + P_N \lambda P_N) G \right ).
\end{equation}
The projection operator $P_N(\textit{\textbf{\textit
r}},\textit{\textbf{\textit r}}^{\prime}) = \sum_k
\phi_{Nk}(\textit{\textbf{\textit r}})
\phi_{Nk}^{*}(\textit{\textbf{\textit r}}^{\prime})$ in the action
\eqref{S1} indicates that the potential
$\lambda(\textit{\textbf{\textit r}})$ exists on the $N$th Landau
level only~\cite{Foot1}. The saddle-point Green function
$G^{\alpha \beta}_{nm}(\textit{\textbf{\textit
r}},\textit{\textbf{\textit r}}^{\prime})$ is determined as
\begin{gather}
G^{\alpha\beta}_{nm}(\textit{\textbf{\textit
r}},\textit{\textbf{\textit
r}}^{\prime})=\delta^{\alpha\beta}\delta_{nm}
\sum\limits_{pk} \phi_{pk}^{*}(\textit{\textbf{\textit r}})\phi_{pk}(\textit{\textbf{\textit r}}^{\prime})G^{-1}_p(\omega_n), \notag\\
G^{-1}_p(\omega_n)=i \omega_n + \omega_H(p-N)+i \frac{\sgn
\omega_n}{2 \tau},\label{Gsp}
\end{gather}
We notice that the Green function \eqref{Gsp} coincides with the
Green function averaged over disorder in the self-consistent Born
approximation~\cite{AFS}.

\begin{figure}[t]
\centerline{\includegraphics[width=200pt,height=100pt]{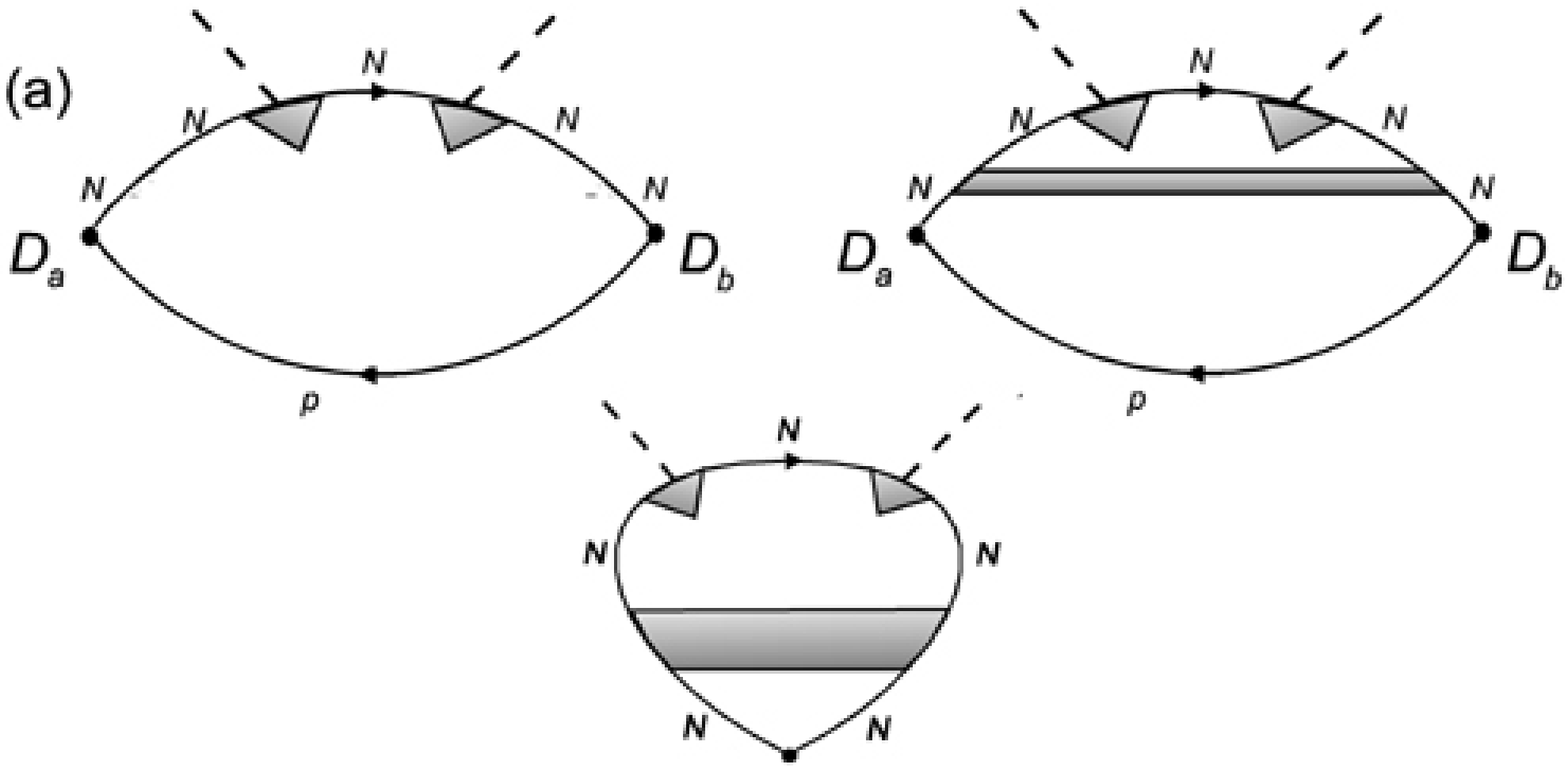}}
\centerline{\includegraphics[width=200pt,height=64pt]{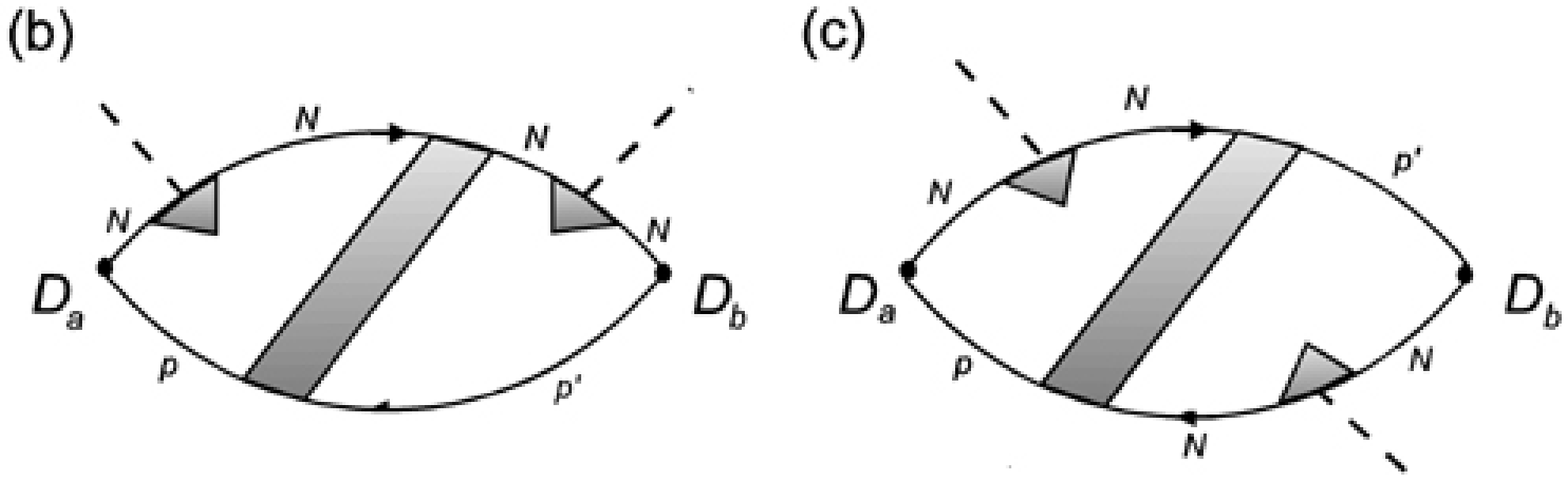}}
\caption{Diagrams for the corrections to $\sigma_{ab}$. The solid
lines are the Green functions, the $N / p / p^{\prime}$ symbols
denote the Landau level, the dashed lines are the UCDW induced
potential $\lambda(\textit{\textbf{\textit r}})$ and the shaded
blocks are impurity ladders.} \label{FIG4}
\end{figure}

%
\textbf{6. Conductivity tensor.} With the action \eqref{S1} in
hands we can evaluate the contributions to the conductivity tensor
$\sigma_{ab}$ due to the presence of the UCDW state on the
half-filled Landau level. As one can verify the contributions of
the first order in the UCDW induced potential
$\lambda(\textit{\textbf{\textit r}})$ vanish. In order to find
the contributions to $\sigma_{ab}$ of the second order in
$\lambda$ we should expand the action \eqref{S1} upto the second
order both in $\lambda$ and $K$. Then, integrating over the $U$
fields, we obtain several contributions. We present the diagrams
that correspond to them in the standard perturbative technique in
Fig.~\ref{FIG4}.

\begin{figure}[t]
\centerline{\includegraphics[width=160pt,height=56pt]{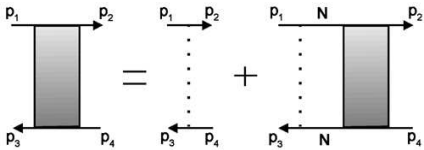}}
\caption{Fig.~\protect\ref{FIG5}. The equation for the impurity
ladder. The frequency $\omega_n+\nu_n$ runs to the right whereas
the $\omega_n$ to the left.} \label{FIG5}
\end{figure}

The first three diagrams (Fig.~\ref{FIG4}(a)) yield only the
isotropic contribution
\begin{gather}
\sigma_{ab}^{(a)}(\nu_n)= - T \sum\limits_{\omega_n}
\frac{G_N^3(\omega_n)}{1+g \pi^{\omega_n}(0,0)}\frac{T_0^2
\Delta^2}{(1+g
\pi^{\omega_n}(0,Q))^2}\notag \\
\times \frac{4\pi\omega_H}{\nu_n}\left (\frac{2}{m}\sum \limits_p
D_{Np}^a D_{pN}^b G_p(\omega_n+\nu_n) - \delta_{ab}\right ).
\label{sigmaA}
\end{gather}
Here the polarization operator $\pi^{\omega_n}(\nu_n,q)$ on the
$N$th Landau level is defined as
\begin{equation}\label{polop}
\pi^{\omega_n}(\nu_n,q) = - n_L G_N(\omega_n+\nu_n) G_N(\omega_n)
F^2_{NN}(q).
\end{equation}
The contribution of diagram Fig.~\ref{FIG4}(b) is seemed to be
proportional to $\mathcal{J}_0(r_0)$ and vanishes therefore. The
last diagram Fig.~\ref{FIG4}(c) is as follows
\begin{gather}
\sigma_{ab}^{(c)}(\nu_n)= \frac{8\pi\omega_H}{\nu_n m}
 T \sum\limits_{\omega_n}
\frac{G_N^2(\omega_n)G^2_N(\omega_n+\nu_n)T_0^2\Delta^2}{(1+g
\pi^{\omega_n}(0,Q))^2}\notag\\\times\sum \limits_{p p^{\prime}}
G_p(\omega_n)G_{p^{\prime}}(\omega_n+\nu_n) \frac{D_{pN}^a
D_{p^{\prime}N}^bI_{NpNp^{\prime}}(\textit{\textbf{\textit
Q}})}{1+g \pi^{\omega_n}(0,Q)}, \label{sigmaC}
\end{gather}
where a single impurity line is written in the Landau level
indices representation (see Fig.~\ref{FIG5}) as
\begin{equation}\label{I}
I_{p_1p_2p_3p_4}(\textit{\textbf{\textit Q}}) = g \int_q
F_{p_1p_2}(\textit{\textbf{\textit q}})
F_{p_3p_4}(-\textit{\textbf{\textit q}}) e^{-i
\textit{\textbf{\textit q}} \textit{\textbf{\textit Q}} l_H^2}.
\end{equation}
The contribution \eqref{sigmaC} contains the anisotropic as well
as isotropic corrections to $\sigma_{ab}$. If we take
$p=p^{\prime}=N\pm 1$ we obtain the anisotropic contribution due
to the structure of matrix elements $D_{p_1p_2}^a$. The opposite
case $p=N\pm 1$ and $p^{\prime}=N\pm 1$ results in the isotropic
correction.


Now with a help of the identities $I_{N,N\pm 1,N,N\pm 1}= g e^{\mp
2 i \phi} \mathcal{J}_1^2(r_0)$ and $I_{N,N\pm 1,N,N\mp 1}= g
\mathcal{J}_1^2(r_0)$ we perform the summation over the Landau
level indices as well as over the Matsubara frequency. As the last
step, we express the UCDW order parameter $\Delta$ via the
temperature difference $T_c-T\ll T_c$ as~\cite{Burm1}
\begin{equation}\label{D2}
\Delta = \zeta\left (2,\frac12+\eta_c\right
)\sqrt{\mathcal{G}\left (\eta_c\right )}\sqrt{\frac{T_c-T}{T_c}}
\end{equation}
and obtain the results \eqref{cond1},\eqref{cond2} and
\eqref{cond3}.

%

\textbf{7. Fluctuations of the order parameter.} The UCDW order
parameter $\Delta$ involved in Eq.\eqref{S1} can be thought of as
a saddle-point solution for the plasmon field that appears in the
Hubbard-Stratonovich transformation of the screened
electron-electron interaction in the action \eqref{Sinit}. The
expansions of physical quantities like free energy and linear
response in $\Delta$ is legitimate if we can neglect the
fluctuations of the UCDW order parameter $\Delta$. It was shown
that they leads to the first order transition at lower temperature
$T_c-\delta T$ where $\delta T/T_c \propto N^{-2/3}$~\cite{Burm1}.
Therefore, in the considered case of the weak magnetic field ($N
\gg 1)$ the effect of the fluctuations on the transition is
negligible and the mean-field picture is well justified. There is
a legitimate question about the effect of the fluctuations on the
conductivity tensor above and below the transition temperature
$T_c$. Below we consider the former case as more interesting.

At $T>T_c$ the mean-field order parameter $\Delta=0$ in average,
but the average of its square $\langle \Delta^2\rangle$ is
non-zero. It leads to the appearance of the corrections to the
conductivity tensor $\sigma_{ab}$ above $T_c$ due to the presence
of the CDW correlations. We can find the contributions of the
order parameter fluctuations to $\sigma_{ab}$ by substituting
$\langle \Delta^2\rangle$ for $\Delta^2$ in Eqs.\eqref{sigmaA} and
\eqref{sigmaC}.

Generally, the angle $\phi$ and modulus $Q$ of the CDW vector
$\textit{\textbf{\textit Q}}$ can fluctuate
simultaneously~\cite{foot2}. Naturally, only the isotropic
correction can appear in this case. Then, the result for the
correction one can obtain from Eq.\eqref{cond3} with a help of the
following substitutions
\begin{gather}
\frac{T_c-T}{T_c} \to \frac{r_0}{4 \pi
N}\sqrt{\frac{T_c}{T-T_c}},\label{TcC}\\
\mathcal{G}\left (z\right )\to
\sqrt{\frac{\zeta^{-1}\left(2,\frac12+z\right)}{\gamma
\zeta\left(2,\frac12+z\right)+ \mathcal{J}^2_1(r_0)z^2
\zeta\left(4,\frac12+z\right)}},\notag
\end{gather}
where $\gamma = \partial \ln T_0/\partial r_0\approx 2.58$. It is
worthwhile to mention that these fluctuational contribution
\eqref{TcC} to $\sigma_{ab}$ above $T_c$ is analogous to the
correction for conductivity of a normal metal due to
superconducting pairing~\cite{AL}. The fluctuation correction
\eqref{TcC} has square-root divergence at $T\to T_c$. This fact
indicates that the result \eqref{TcC} is not applicable in the
vicinity of the transition temperature $T_c$. The limit of
applicability is determined by the requirement that the
fluctuational correction should be much smaller than $\sigma_{ab}$
itself.

%
\textbf{8. Conclusion.} Summarizing, we calculated the anisotropic
as well as isotropic corrections to the conductivity tensor of the
two dimensional electrons on the half-filled high Landau level
just below the transition to the UCDW state. The corrections
obtained are proportional to $(T_c-T)/T_c$ that is in agreement
with that found in the experiments. Also we calculated the
fluctuational correction to the conductivity tensor of the
two-dimensional electrons above the transition.

%
I am grateful to M.A. Baranov, L.I. Glazman, M.V. Feigelman, P.M.
Ostrovsky, M.A. Skvortsov for illuminating discussions. Financial
support from Russian Foundation for Basic Research (RFBR), the
Russian Ministry of Science, Forschungszentrum J\"ulich (Landau
Scholarship), and Dutch Science Foundation (FOM) is acknowledged.



\begin{thebibliography}{99}
\bibitem{Foot1} In general, the UCDW state on the $N$th Landau level creates the
potential not only on the same level but also on the adjacent
ones. We neglect this effect since it results in the corrections
to the conductivity tensor $\sigma_{ab}$ of the order of $1/N^2$
beside expressions \eqref{cond1},\eqref{cond2}, and \eqref{cond3}.
\end{thebibliography}
\end{document}